\documentclass{llncs}
\usepackage{version}
\usepackage{pasis}

\title{\mbox{} \\*[3.5ex]
Process Algebra, Process Scheduling, \\ and Mutual Exclusion}
\author{C.A. Middelburg}
\institute{Informatics Institute, Faculty of Science, University of
           Amsterdam \\
           Science Park~904, 1098~XH Amsterdam, the Netherlands \\
           \email{C.A.Middelburg@uva.nl}}

\begin{document}
\maketitle

\begin{abstract}
%% 89 %%
In the case of multi-threading as found in contemporary programming 
languages, parallel processes are interleaved according to what is known 
as a process-scheduling policy in the field of operating systems.
In a previous paper, we extend \ACP\ with this form of interleaving.
In the current paper, we do so with the variant of \ACP\ known as \ACPe. 
The choice of \ACPe\ stems from the need to cover more 
process-scheduling policies. 
We show that a process-scheduling policy supporting mutual exclusion 
of critical subprocesses is now covered.
\begin{classcode}
D.1.3, D.4.1, F.1.2
\end{classcode}
\end{abstract}

\section{Introduction}
\label{sect-intro}

In algebraic theories of processes, such as \ACP~\cite{BW90}, 
CCS~\cite{Mil89}, and CSP~\cite{Hoa85}, processes are discrete 
behaviours that proceed by doing steps in a sequential fashion.
In these theories, parallel composition of two processes is usually
interpreted as arbitrary interleaving of the steps of the processes 
concerned. 
Arbitrary interleaving turns out to be appropriate for many applications 
and to facilitate formal algebraic reasoning. 
Our interest in process-scheduling policies originates from the feature 
of multi-threading found in contemporary programming languages such as 
Java~\cite{GJSB00a} and C\#~\cite{HWG03a}.
Multi-threading gives rise to parallel composition of processes.
However, the steps of the processes concerned are interleaved according 
to what is known as a process-scheduling policy in the field of 
operating systems.
We use the term strategic interleaving for this more constrained form of
interleaving and the term interleaving strategy instead of 
process-scheduling policy.

Nowadays, multi-threading is often used in the implementation of 
systems.
Because of this, in many systems, for instance hardware/software 
systems, we have to do with parallel processes that may best be 
considered to be interleaved in an arbitrary way as well as parallel 
processes that may best be considered to be interleaved according to 
some interleaving strategy.
To our knowledge, there exists no work on strategic interleaving in the 
setting of a general algebraic theory of processes like ACP, CCS and 
CSP.
This is what motivated us to do the work presented in~\cite{BM17b}, 
namely extending \ACP\ such that it supports both arbitrary interleaving
and strategic interleaving.

The extension of \ACP\ presented in~\cite{BM17b} is based on a generic 
interleaving strategy that can be instantiated with different specific
interleaving strategies.
The main reason for doing the work presented in the current paper is the 
finding that the generic interleaving strategy concerned cannot be 
instantiated with:
(a)~interleaving strategies where the data relevant to the 
process-scheduling decision making may be such that none of the 
processes concerned can be given a turn,
(b)~interleaving strategies where the data relevant to the 
process-scheduling decision making must be updated on successful 
termination of one of the processes concerned, and
(c)~interleaving strategies where the process-scheduling decision making 
may be adjusted by steps of the processes concerned that are solely 
intended to change the data relevant to the process-scheduling decision 
making.
Another reason is the fact that it is not shown in~\cite{BM17b} that the 
generic interleaving strategy can be instantiated with non-trivial 
specific interleaving strategies.

In this paper, we rectify the shortcomings of the generic interleaving 
strategy on which the extension of \ACP\ presented in~\cite{BM17b} is 
based by starting from \ACPe, i.e.\ \ACP\ extended with a constant for a 
process that can only terminate successfully, and widening the generic 
interleaving strategy.
Moreover, it is shown that the widened generic interleaving strategy can
be instantiated with an interleaving strategy that supports mutual 
exclusion of critical subprocesses of the different processes being 
interleaved.

The relevance of developing theory about strategic interleaving lies in 
the fact that strategic interleaving is quite different from arbitrary
interleaving.
This is for example illustrated by the following easy to demonstrate 
phenomena: 
(a)~sometimes a particular interleaving strategy leads to inactiveness 
whereas arbitrary interleaving does not lead to inactiveness and 
(b)~whether the interleaving of certain processes leads to inactiveness 
depends on the interleaving strategy\linebreak[2] used.

The rest of this paper is organized as follows.
First, we review \ACPe\ (Section~\ref{subsect-ACPe}) and guarded 
recursion in the setting of \ACPe\ (Section~\ref{subsect-ACPer}).
Then, we extend \ACPe\ with strategic interleaving 
(Section~\ref{subsect-SI}) and present some properties concerning the
connection between \ACPe\ and this extension 
(Section~\ref{subsect-theorems-siACP}).  
After that, we show that the generic interleaving strategy on which the 
extension is based can be instantiated with the interleaving strategy 
mentioned above (Section~\ref{sect-mutex}).
Finally, we make some concluding remarks (Section~\ref{sect-concl}).

In~\cite{BM17b}, \cite{Mid19a}, and the current paper, different 
variants of \ACP\ are extended with strategic interleaving.
Because of this, there is some text overlap between these papers.
The current paper can be looked upon as supplementary material 
to~\cite{BM17b}, but can be read independently.

\section{\ACPe\ with Guarded Recursion}
\label{sect-ACPer}

In this section, we give a survey of \ACPe\ (\ACP\ with the empty 
process) and guarded recursion in the setting of \ACPe.
For a more comprehensive treatment, the reader is referred 
to~\cite{BW90}.

\subsection{\ACPe}
\label{subsect-ACPe}

In \ACPe, it is assumed that a fixed but arbitrary set $\Act$ of
\emph{actions}, with $\dead, \ep \notin \Act$, has been given.
We write $\Actd$ for $\Act \union \set{\dead}$ and
$\Acte$ for $\Act \union \set{\ep}$.
It is further assumed that a fixed but arbitrary commutative and 
associative \emph{communication} function 
$\funct{\commf}{\Actd \x \Actd}{\Actd}$, with 
$\commf(\dead,a) = \dead$ for all $a \in \Actd$, has been given.
The function $\commf$ is regarded to give the result of synchronously
performing any two actions for which this is possible, and to give 
$\dead$ otherwise.

The signature of \ACPe\ consists of the following constants and 
operators:
\begin{itemize}
\item
the \emph{inaction} constant $\dead$\,;
\item
the \emph{empty process} constant $\ep$\,;
\item
for each $a \in \Act$, the \emph{action} constant 
$a$\,;
\item
the binary \emph{alternative composition} operator 
$\ph \altc \ph$\,;
\item
the binary \emph{sequential composition} operator 
$\ph \seqc \ph$\,;
\item
the binary \emph{parallel composition} operator 
$\ph \parc \ph$\,;
\item
the binary \emph{left merge} operator 
$\ph \leftm \ph$\,;
\item
the binary \emph{communication merge} operator 
$\ph \commm \ph$\,;
\item
for each $H \subseteq \Act$, the unary \emph{encapsulation} operator
$\encap{H}$\,.
\end{itemize}
We assume that there is a countably infinite set $\cX$ of variables 
which contains $x$, $y$ and $z$ with and without subscripts.
Terms are built as usual.
We use infix notation for the binary operators.
The precedence conventions used with respect to the operators of \ACPe\
are as follows: $\altc$ binds weaker than all others, $\seqc$ binds
stronger than all others, and the remaining operators bind equally
strong.

The constants of \ACPe\ can be explained as follows ($a \in \Act$):
\begin{itemize}
\item
$\dead$ denotes the process that cannot do anything;
\item
$\ep$ denotes the process that can only terminate successfully;
\item
$a$ denotes the process that can first perform action $a$ and after that 
terminate successfully.
\end{itemize}
Let $t$ and $t'$ be closed \ACPe\ terms denoting processes $p$ and $p'$.
Then the operators of \ACPe\ can be explained as follows:
\begin{itemize}
\item
$t \altc t'$ denotes the process that can behave as $p$ or as $p'$, but 
not both;
\item
$t \seqc t'$\, denotes the process that can first behave as $p$ and 
after that as $p'$;
\item
$t \parc t'$ denotes the process that can behave as $p$ and $p'$ in 
parallel;
\item
$t \leftm t'$ denotes the same process as $t \parc t'$, except that it 
must start with performing an action of $p$;
\item
$t \commm t'$ denotes the same process as $t \parc t'$, except that it 
must start with \mbox{performing} an action of $p$ and an action of $p'$ 
synchronously;
\item
$\encap{H}(t)$ denotes the process that can behave the same as $p$, 
except that actions from $H$ are blocked.
\end{itemize}
The operators $\leftm$ and $\commm$ are of an auxiliary nature.
They are needed to axiomatize \ACPe.

The axioms of \ACPe\ are the equations given in Table~\ref{axioms-ACPe}.
\begin{table}[!t]
\caption{Axioms of \ACPe}
\label{axioms-ACPe}
\begin{eqntbl}
\begin{axcol}
x \altc y = y \altc x                                  & \axiom{A1}   \\
(x \altc y) \altc z = x \altc (y \altc z)              & \axiom{A2}   \\
x \altc x = x                                          & \axiom{A3}   \\
(x \altc y) \seqc z = x \seqc z \altc y \seqc z        & \axiom{A4}   \\
(x \seqc y) \seqc z = x \seqc (y \seqc z)              & \axiom{A5}   \\
x \altc \dead = x                                      & \axiom{A6}   \\
\dead \seqc x = \dead                                  & \axiom{A7}   \\
x \seqc \ep = x                                        & \axiom{A8}   \\
\ep \seqc x = x                                        & \axiom{A9}   \\
{}                                                                    \\
\encap{H}(\ep) = \ep                                   & \axiom{D0}   \\
\encap{H}(a) = a                \hfill \mif a \notin H & \axiom{D1}   \\
\encap{H}(a) = \dead            \hfill \mif a \in H    & \axiom{D2}   \\
\encap{H}(x \altc y) = \encap{H}(x) \altc \encap{H}(y) & \axiom{D3}   \\
\encap{H}(x \seqc y) = \encap{H}(x) \seqc \encap{H}(y) & \axiom{D4}  
\end{axcol}
\qquad
\begin{axcol}
x \parc y = {} \\ \quad
x \leftm y \altc y \leftm x \altc x \commm y \altc 
\encap{\Act}(x) \seqc \encap{\Act}(y)                  & \axiom{CM1T} \\
\ep \leftm x = \dead                                   & \axiom{CM2T} \\
a \seqc x \leftm y = a \seqc (x \parc y)               & \axiom{CM3}  \\
(x \altc y) \leftm z = x \leftm z \altc y \leftm z     & \axiom{CM4}  \\
\ep \commm x = \dead                                   & \axiom{CM5T} \\
x \commm \ep = \dead                                   & \axiom{CM6T} \\
a \seqc x \commm b \seqc y =
                        \commf(a,b) \seqc (x \parc y)  & \axiom{CM7}  \\
(x \altc y) \commm z = x \commm z \altc y \commm z     & \axiom{CM8}  \\
x \commm (y \altc z) = x \commm y \altc x \commm z     & \axiom{CM9}  \\
a \commm b = \commf(a,b)                               & \axiom{CM10} \\
\dead \commm x = \dead                                 & \axiom{CM11} \\
x \commm \dead = \dead                                 & \axiom{CM12} 
\end{axcol}
\end{eqntbl}
\end{table}
In these equations, $a$ and $b$ stand for arbitrary constants of \ACPe\ 
that differ from $\ep$, and $H$ stands for an arbitrary subset of 
$\Act$.
Moreover, $\commf(a,b)$ stands for the action constant for the action 
$\commf(a,b)$.
In D1 and D2, side conditions restrict what $a$ and $H$ stand for.

In some presentations of \ACPe, e.g.\ in~\cite{BW90},
$\encap{\Act}(x) \seqc \encap{\Act}(y)$
is replaced by $\surd(x) \seqc \surd(y)$ in CM1T. 
However, $\encap{\Act}$ and $\surd$ have the same axioms.
In other presentations of \ACPe, $\commf(a,b)$ is frequently replaced by 
$a \commm b$ in CM7.
By CM10, which is more often called CF, this replacement give rise to an 
equivalent axiomatization.
In other presentations of \ACPe, CM11 and CM12 are usually absent.
These equations are not derivable from the other axioms, but all there 
closed substitution instances are derivable from the other axioms.
Moreover, CM11 and CM12 hold in virtually all models of \ACPe\ that have 
been devised.

\subsection{Guarded Recursion}
\label{subsect-ACPer}

A closed \ACPe\ term denotes a process with a finite upper bound to the 
number of actions that it can perform. 
Guarded recursion allows the description of processes without a finite 
upper bound to the number of actions that it can perform.

This section applies to both \ACPe\ and its extension \siACPe\ 
introduced in Section~\ref{sect-SI}.  
Therefore, in this section, let \APA\ be \ACPe\ or \siACPe. 

Let $t$ be a \APA\ term containing a variable $X$.
Then an occurrence of $X$ in $t$ is \emph{guarded} if $t$ has a subterm 
of the form $a \seqc t'$ where $a \in \Act$ and $t'$ is a \APA\ term 
containing this occurrence of $X$.
A \APA\ term $t$ is a \emph{guarded} \APA\ term if all occurrences of 
variables in $t$ are guarded.

A \emph{guarded recursive specification} over \APA\ is a set 
$\set{X_i = t_i \where i \in I}$, 
where $I$ is finite or countably infinite set, 
each $X_i$ is a variable from $\cX$, 
each $t_i$ is either a guarded \APA\ term in which variables other than 
the variables from $\set{X_i \where i \in I}$ do not occur or a \APA\ 
term rewritable to such a term using the axioms of \APA\ in either 
direction and/or the equations in 
$\set{X_j = t_j \where j \in I \Land i \neq j}$ 
from left to right, and $X_i \neq X_j$ for all $i,j \in I$ with 
$i \neq j$.

We write $\vars(E)$, where $E$ is a guarded recursive specification, for 
the set of all variables that occur in $E$.

A solution of a guarded recursive specification $E$ in some model of 
\APA\ is a set $\set{p_X \where X \in \vars(E)}$ of elements of the 
carrier of that model such that each equation in $E$ holds if, for all 
$X \in \vars(E)$, $X$ is assigned $p_X$.
We are only interested in models of \APA\ in which guarded recursive 
specifications have unique solutions. 

We extend \APA\ with guarded recursion by adding constants for solutions 
of guarded recursive specifications over \APA\ and axioms concerning 
these additional constants.
For each guarded recursive specification $E$ over \APA\ and each 
$X \in \vars(E)$, we add a constant, denoted by $\rec{X}{E}$, that 
stands for the unique solution of $E$ for $X$ to the constants of \APA.
We add the equation RDP and the conditional equation RSP given in 
Table~\ref{axioms-REC} to the axioms of \APA.
\begin{table}[!t]
\caption{Axioms for guarded recursion}
\label{axioms-REC}
\begin{eqntbl}
\begin{saxcol}
\rec{X}{E} = \rec{t}{E} & \mif X = t\; \in \;E          & \axiom{RDP} \\
E \Limpl X = \rec{X}{E}   & \mif X \in \vars(E)         & \axiom{RSP} 
\end{saxcol}
\end{eqntbl}
\end{table}
In RDP and RSP, $X$ stands for an arbitrary variable from $\cX$, $t$ 
stands for an arbitrary \APA\ term, $E$ stands for an arbitrary 
guarded recursive specification over \APA, and the notation $\rec{t}{E}$ 
is used for $t$ with, for all $X \in \vars(E)$, all occurrences of $X$ 
in $t$ replaced by $\rec{X}{E}$.
Side conditions restrict what $X$, $t$ and $E$ stand for.
We write $\APA_\mathrm{rec}$ for the resulting theory.

The equations $\rec{X}{E} = \rec{t}{E}$ for a fixed $E$ express that 
the constants $\rec{X}{E}$ make up a solution of $E$ and 
the conditional equations $E \Limpl X = \rec{X}{E}$ express that this 
solution is the only one.

In extensions of \ACPe\ whose axioms include RSP, we have to deal with 
conditional equational formulas with a countably infinite number of 
premises.
Therefore, infinitary conditional equational logic is used in deriving 
equations from the axioms of extensions of \ACPe\ whose axioms include 
RSP.
A complete inference system for infinitary conditional equational logic 
can be found in, for example, \cite{GV93}.
It is noteworthy that in the case of infinitary conditional equational 
logic derivation trees may be infinitely branching (but they may not 
have infinite branches).

\section{Strategic Interleaving}
\label{sect-SI}

In this section, we extend \ACPe\ with strategic interleaving, i.e.\
interleaving according to some interleaving strategy.
Interleaving strategies are abstractions of scheduling algorithms.
Interleaving according to some interleaving strategy represents what 
really happens in the case of multi-threading as found in contemporary 
programming languages.

\subsection{\ACPe\ with Strategic Interleaving}
\label{subsect-SI}

In the extension of \ACP\ with strategic interleaving presented below, 
it is expected that an interleaving strategy uses the interleaving 
history in one way or another to make process-scheduling decisions.

The set $\Hist$ of \emph{interleaving histories} is the subset of 
$\seqof{(\Natpos \x \Natpos)}$ that is inductively defined by the 
following rules:%
\footnote
{The sequence notation used in this paper is explained in 
 Appendix~\ref{appendix-notations}.}
\begin{itemize}
\item
$\emptyseq \in \Hist$;
\item
if $i \leq n$, then $\tup{i,n} \in \Hist$;
\item
if $h \concat \tup{i,n} \in \Hist$, $j \leq n$, and 
$n - 1 \leq m \leq n + 1$, then 
$h \concat \tup{i,n} \concat \tup{j,m} \in \Hist$.
\end{itemize}
The intuition concerning interleaving histories is as follows:
if the $k$th pair of an interleaving history is $\tup{i,n}$, then the 
$i$th process got a turn in the $k$th interleaving step and after its
turn there were $n$ processes to be interleaved.
The number of processes to be interleaved may increase due to process
creation (introduced below) and decrease due to successful termination 
of processes.
 
The presented extension of \ACPe\ is called \siACPe\ (\ACP\ with 
Strategic Interleaving). 
It is based on a generic interleaving strategy that can be instantiated 
with different specific interleaving strategies that can be represented 
in the way that is explained below.

In \siACPe, it is assumed that the following has been given:% 
\footnote
{We write $\pfunct{f}{A}{B}$ to indicate that $f$ is a partial function
 from $A$ to $B$.}
\begin{itemize}
\item
a fixed but arbitrary set $S$; 
\item
a fixed but arbitrary partial function 
$\pfunct{\sched{n}}{\Hist \x S}{\set{1,\ldots,n}}$ 
for each $n \in \Natpos$;
\item
a fixed but arbitrary total function 
$\funct{\updat{n}}{\Hist \x S \x \set{1,\ldots,n} \x \Acte}{S}$ 
for each $n \in \Natpos$;
\item
a fixed but arbitrary set $C \subset \Act$ such that, 
for each $c \in C$, $\ol{c} \in \Act \diff C$ and, 
for each $a,b \in \Act$, 
$\commf(a,b) \neq c$, $\commf(a,b) \neq \ol{c}$,
$\commf(a,c) = \dead$, and $\commf(a,\ol{c}) = \dead$.
\end{itemize}
The elements of $S$ are called \emph{control states}, $\sched{n}$ is 
called an \emph{abstract scheduler} (\emph{for $n$ processes}),  
$\updat{n}$ is called a \emph{control state transformer} 
(\emph{for $n$ processes}), and 
the elements of $C$ are called \emph{control actions}.
The intuition concerning $S$, $\sched{n}$, $\updat{n}$, and $C$ is as 
follows:
\begin{itemize}
\item
the control states from $S$ encode data that are relevant to the 
interleaving strategy, but not derivable from the interleaving history;
\item
if $\sched{n}(h,s) = i$, then the $i$th process gets the next turn after 
interleaving history $h$ in control state $s$;
\item
if $\sched{n}(h,s)$ is undefined, then no process gets the next turn 
after interleaving history $h$ in control state $s$; 
\item
if $\updat{n}(h,s,i,a) = s'$, then $s'$ is the control state that arises 
from the $i$th process\linebreak[2] doing $a$ after interleaving history 
$h$ in control state $s$;
\item
if $\updat{n}(h,s,i,\ep) = s'$, then $s'$ is the control state that 
arises from the $i$th process terminating successfully after 
interleaving history $h$ in control state~$s$;
\item
if $a \in C$, then $a$ is an explicit means to bring about a control 
state change and $\ol{a}$ is left as a trace after $a$ has been dealt 
with. 
\end{itemize}
Thus, $S$, $\indfam{\sched{n}}{n \in \Natpos}$,  
$\indfam{\updat{n}}{n \in \Natpos}$, and $C$ make up a way to represent 
an interleaving strategy.
This way to represent an interleaving strategy is engrafted 
on~\cite{SS00a}.

The intuition concerning the actions $\ol{a}$, where $a \in C$, is as 
follows: when a process performs a control action $a$, it will interact 
with the scheduler at hand and the action resulting from this 
interaction is the action $\ol{a}$.

In \siACP, the extension of \ACP\ with strategic interleaving 
from~\cite{BM17b}, abstract schedulers must be total functions
$\funct{\sched{n}}{\Hist \x S}{\set{1,\ldots,n}}$, control state 
transformers must be total functions 
$\funct{\updat{n}}{\Hist \x S \x \set{1,\ldots,n} \x \Act}{S}$, and
control actions are not distinguished from other actions. 
The widenings chosen in \siACPe\ rectify the shortcomings mentioned 
in Section~\ref{sect-intro}.
The widening with respect to the control state transformers would not
be possible without the change from \ACP\ to \ACPe.

Consider the case where $S$ is a singleton set, 
for each $n \in \Natpos$, $\sched{n}$ is defined by
\begin{ldispl}
\sched{n}(\emptyseq,s) = 1\;,  \\
\sched{n}(h \concat \tup{j,n},s) = (j \bmod n) + 1\;,
\end{ldispl}%
for each $n \in \Natpos$, $\updat{n}$ is defined by 
\begin{ldispl}
\updat{n}(h,s,i,a) = s\;,
\end{ldispl}%
and $C$ is the empty set.
In this case, the interleaving strategy corresponds to the round-robin 
scheduling algorithm for the case where each process is given only one 
turn in a row.
More advanced strategies can be obtained if the scheduling makes more 
advanced use of the interleaving history and the control state.
An example is given in Section~\ref{sect-mutex}.

In \siACPe, it is also assumed that a fixed but arbitrary finite or
countably infinite set $D$ of \emph{data} and a fixed but arbitrary 
function $\funct{\crea}{D}{P}$, where $P$ is the set of all closed terms 
over the signature of \siACPe\ (given below), have been given and that, 
for each $d \in D$, 
$\pcr(d),\rcr(d) \in
 \Act \diff (C \union \set{\ol{c} \where i \in C})$ and, 
for each $a, b \in \Act$,
$\commf(a,b) \neq \pcr(d)$, $\commf(a,b) \neq \rcr(d)$,
$\commf(a,\pcr(d)) = \dead$, and $\commf(a,\rcr(d)) = \dead$.
The action $\pcr(d)$ can be considered a process creation request and 
the action $\rcr(d)$ can be considered a process creation act.
They stand for the request to start the process denoted by $\crea(d)$ in 
parallel with the requesting process and the act of carrying out that 
request, respectively.

The signature of \siACPe\ consists of the constants and operators
from the signature of \ACP\ and in addition the following operators:
\begin{itemize}
\item
the $n$-ary \emph{strategic interleaving} operator $\siop{n}{h}{s}$
for each $n \in \Natpos$, $h \in \Hist$, and $s \in S$;
\item
the $n$-ary \emph{positional strategic interleaving} operator
$\posmop{n}{i}{h}{s}$
for each $n,i \in \Natpos$ with $i \leq n$, $h \in \Hist$, and 
$s \in S$.
\end{itemize}

The strategic interleaving operators can be explained as follows:
\begin{itemize}
\item
a closed term of the form $\si{n}{h}{s}{t_1,\ldots,t_n}$ denotes the 
process that results from interleaving of the $n$ processes denoted by 
$t_1,\ldots,t_n$ after interleaving history $h$ in control state $s$, 
according to the interleaving strategy represented by $S$, 
$\indfam{\sched{n}}{n \in \Natpos}$, $\indfam{\updat{n}}{n \in \Natpos}$,
and $C$.
\end{itemize}
The positional strategic interleaving operators are auxiliary operators 
used to axiomatize the strategic interleaving operators.
The role of the positional strategic interleaving operators in the 
axiomatization is similar to the role of the left merge operator found 
in \ACPe.

The axioms of \siACPe\ are the axioms of \ACPe\ and in addition the 
equations given in Table~\ref{axioms-strategic-interleaving}.%
\footnote
{There is no axiom named SI6 because for common axioms of \siACP\ and 
 \siACPe\ the names introduced in~\cite{BM17b} have been adopted.}
\begin{table}[!t]
\caption{Axioms for strategic interleaving}
\label{axioms-strategic-interleaving}
\begin{eqntbl}
\begin{axcol}
\si{n}{h}{s}{x_1,\ldots,x_n} = \dead  
\hfill \mif \sched{n}(h,s) \mathrm{\;is\;undefined}    & \axiom{SI0} \\ 
\si{n}{h}{s}{x_1,\ldots,x_n} = 
\posm{n}{\sched{n}(h,s)}{h}{s}{x_1,\ldots,x_n}         
\hfill \mif \sched{n}(h,s) \mathrm{\;is\;defined}\phantom{\mathrm{un}}       
                                                       & \axiom{SI1}  
\eqnsep
\posm{n}{i}{h}{s}{x_1,\ldots,x_{i-1},\dead,x_{i+1},\ldots,x_n} = \dead
                                                       & \axiom{SI2} \\
\posm{1}{i}{h}{s}{\ep} = \ep                           & \axiom{SI3T} \\
\posm{n+1}{i}{h}{s}{x_1,\ldots,x_{i-1},\ep,x_{i+1},\ldots,x_{n+1}} =
\\ \qquad
\si{n}{h \concat \tup{i,n}}{\updat{n+1}(h,s,i,\ep)}
 {x_1,\ldots,x_{i-1},x_{i+1},\ldots,x_{n+1}}           & \axiom{SI4T} \\
\posm{n}{i}{h}{s}{x_1,\ldots,x_{i-1},a \seqc x_i',x_{i+1},\ldots,x_n} =
\\ \qquad
a \seqc
\si{n}{h \concat \tup{i,n}}{\updat{n}(h,s,i,a)}
 {x_1,\ldots,x_{i-1},x_i',x_{i+1},\ldots,x_n}
\hfill \mif a \notin C                                    & \axiom{SI5Ta} \\
\posm{n}{i}{h}{s}{x_1,\ldots,x_{i-1},a \seqc x_i',x_{i+1},\ldots,x_n} =
\\ \qquad
\ol{a} \seqc
\si{n}{h \concat \tup{i,n}}{\updat{n}(h,s,i,a)}
 {x_1,\ldots,x_{i-1},x_i',x_{i+1},\ldots,x_n}
\hfill \mif a \in C                                 & \axiom{SI5Tb} \\
\posm{n}{i}{h}{s}
 {x_1,\ldots,x_{i-1},\pcr(d) \seqc x_i',x_{i+1},\ldots,x_n} =
\\ \qquad
\rcr(d) \seqc
\si{n+1}{h \concat \tup{i,n+1}}{\updat{n}(h,s,i,\pcr(d))}
 {x_1,\ldots,x_{i-1},x_i',x_{i+1},\ldots,x_n,\crea(d)} & \axiom{SI7} \\ 
\posm{n}{i}{h}{s}
 {x_1,\ldots,x_{i-1},x_i' \altc x_i'',x_{i+1},\ldots,x_n} =
\\ \qquad
\posm{n}{i}{h}{s}{x_1,\ldots,x_{i-1},x_i',x_{i+1},\ldots,x_n} \altc
\posm{n}{i}{h}{s}{x_1,\ldots,x_{i-1},x_i'',x_{i+1},\ldots,x_n} 
                                                       & \axiom{SI8}
\end{axcol}
\end{eqntbl}
\end{table}
In the additional equations, $n$ and $i$ stand for arbitrary numbers 
from $\Natpos$ with $i \leq n$, $h$ stands for an arbitrary interleaving 
history from $\Hist$, $s$ stands for an arbitrary control state from 
$S$, $a$ stands for an arbitrary action constant that is not of the form 
$\pcr(d)$ or $\rcr(d)$, and $d$ stands for an arbitrary datum $d$ from 
$D$.

Axiom SI2 expresses that, in the event of inactiveness of the process 
whose turn it is, the whole becomes inactive immediately.
A plausible alternative is that, in the event of inactiveness of the 
process whose turn it is, the whole becomes inactive only after all 
other processes have terminated or become inactive.
In that case, the functions 
$\funct{\updat{n}}{\Hist \x S \x \set{1,\ldots,n} \x \Acte}{S}$ 
must be extended to functions 
$\funct{\updat{n}}
 {\Hist \x S \x \set{1,\ldots,n} \x (\Acte \union \set{\dead})}{S}$
and axiom SI2 must be replaced by the axioms in 
Table~\ref{axioms-alt-deadlock}.
\begin{table}[!t]
\caption{Alternative axioms for SI2}
\label{axioms-alt-deadlock}
\begin{eqntbl}
\begin{axcol}
\posm{1}{i}{h}{s}{\dead} = \dead                       & \axiom{SI2a} \\
\posm{n+1}{i}{h}{s}{x_1,\ldots,x_{i-1},\dead,x_{i+1},\ldots,x_{n+1}} = 
\\ \qquad
\si{n}{h \concat \tup{i,n}}{\updat{n+1}(h,s,i,\dead)}
      {x_1,\ldots,x_{i-1},x_{i+1},\ldots,x_{n+1}} \seqc \dead         
                                                       & \axiom{SI2b} 
\end{axcol}
\end{eqntbl}
\end{table}

In \siACPer, i.e.\ \siACPe\ extended with guarded recursion in the way 
described in Section~\ref{subsect-ACPer}, the processes that can be 
created are restricted to the ones denotable by a closed \siACPe\ term.
This restriction stems from the requirement that $\crea$ is a function 
from $D$ to the set of all closed \siACPe\ terms.
The restriction can be removed by relaxing this requirement to the 
requirement that $\crea$ is a function from $D$ to the set of all closed 
\siACPer\ terms. 
We write \siACPerp\ for the theory resulting from this relaxation.
In other words, \siACPerp\ differs from \siACPer\ in that it is assumed 
that a fixed but arbitrary function $\funct{\crea}{D}{P}$, where $P$ is 
the set of all closed terms over the signature of \siACPer, has been 
given.

It is customary to associate transition systems with closed terms of the 
language of an ACP-like algebraic theory of processes by means of 
structural operational semantics and to use this to construct a model in 
which closed terms are identified if their associated transition systems 
are bisimilar.
The structural operational semantics of \ACPe\ can be found 
in~\cite{BW90}.
The additional transition rules for the strategic interleaving operators 
and the positional strategic interleaving operators are given in 
Appendix~\ref{appendix-SOS}.

\subsection{On the Connection between \ACPe\ and \siACPe}
\label{subsect-theorems-siACP}

In this section, we present some theorems concerning the connection 
between \ACPe\ and \siACPe.
Each of the theorems refers to more than one process algebra.
It is implicit that the same set $\Act$ of actions and the same 
communication function $\commf$ are assumed in the process algebras
referred to.

Each guarded recursive specification over \siACPe\ can be reduced to a
guarded recursive specification over \ACPe. 
\begin{theorem}[Reduction]
\label{theorem-reduction}
\sloppy
For each guarded recursive specification $E$ over \siACPe\ and each 
$X \in \vars(E)$, there exists a guarded recursive specification $E'$ 
over \ACPe\ such that $\rec{X}{E} = \rec{X}{E'}$ is derivable from the 
axioms of \siACPer.
\end{theorem}

Each closed \siACPe\ term is derivably equal to a closed \ACPe\ term.
\begin{theorem}[Elimination]
\label{theorem-elimination}
% !!
\vspace*{-1ex}
\begin{enumerate}
\item[\textup{\bf 1.}]
For each closed \siACPe\ term $t$, there exists a closed \ACPe\ term 
$t'$ such that $t = t'$ is derivable from the axioms of \siACPe.
\item[\textup{\bf 2.}]
For each closed \siACPerp\ term $t$, there exists a closed 
\ACPer\ term $t'$ such that $t = t'$ is derivable from the axioms of 
\siACPerp.
\end{enumerate}
\end{theorem}

Each equation between closed \ACPe\ terms that is derivable in \siACPe\ is 
also derivable in \ACPe.
\begin{theorem}[Conservative extension]
\label{theorem-conservativity}
For each two closed \ACPe\ terms $t$ and $t'$, $t = t'$ is derivable from 
the axioms of \siACPe\ only if $t = t'$ is derivable from the axioms of 
\ACPe.
\end{theorem}

The following theorem concerns the expansion of minimal models of \ACPe\ 
to models of \siACPe.
\begin{theorem}[Unique expansion]
\label{theorem-expansion}
\sloppy
Each minimal model of \ACPe\ has a unique expansion to a model of 
\siACPe.
\end{theorem}

The proofs of Theorems~\ref{theorem-reduction}, 
\ref{theorem-elimination}.2, \ref{theorem-conservativity}, 
and~\ref{theorem-expansion} go along the same line as the proofs of 
Theorems~1, 2, 3, and~4, respectively, in~\cite{BM20a}.%
\footnote
{The proof outline of Theorem~1 in~\cite{BM20a} is an improvement of the 
inadequate proof outline of Theorem~1 in~\cite{BM17b}.}
Theorem~\ref{theorem-elimination}.1 is a corollary of the proof of
Theorem~\ref{theorem-elimination}.2.
Theorems~\ref{theorem-reduction} and~\ref{theorem-elimination}.2 would 
not go through if guarded recursive specifications were required to be 
finite.

\section{An Example}
\label{sect-mutex}

In this section, we instantiate the generic interleaving strategy on 
which \siACP\ is based with a specific interleaving strategy.
The interleaving strategy concerned corresponds to the round-robin 
scheduling algorithm, where each of the processes being interleaved is 
given a fixed number $k$ of consecutive turns, adapted to mutual 
exclusion of critical subprocesses of the different processes being 
interleaved.
\linebreak[2]
Mutual exclusion of certain subprocesses is the condition that they are 
not interleaved and critical subprocesses are subprocesses that possibly 
interfere with each other when this condition is not met.
The adopted mechanism for mutual exclusion is essentially a binary 
semaphore mechanism~\cite{Dij68a,Bri73a,Ben06a}.
Below binary semaphores are simply called \emph{semaphores}.

In this section, it is assumed that a fixed but arbitrary natural number 
$k \in \Natpos$ has been given.
We use $k$ as the number of consecutive turns that each process being
interleaved gets.

Moreover, it is assumed that a finite set $R$ of semaphores has been 
given.
\linebreak[2]
We instantiate the set $C$ of control actions as follows: 
\begin{ldispl}
C = \set{\Pop(r) \where r \in R} \union \set{\Vop(r) \where r \in R}\;,
\end{ldispl}%
hereby taking for granted that $C$ satisfies the necessary conditions. 
The $\Pop$ and $\Vop$ actions correspond to the $\mathsf{P}$ and 
$\mathsf{V}$ operations from~\cite{Dij68a}.

We instantiate the set $S$ of control states as follows: 
\begin{ldispl}
S = \Union{R' \subseteq R} (\mapof{R'}{\seqof{\Natpos\!}})\;.
\end{ldispl}%
The intuition concerning the connection between control states $s \in S$ 
and the semaphore mechanism as introduced in~\cite{Dij68a} is as 
follows:
\begin{itemize}
\item
$r \notin \dom(s)$ indicates that semaphore $r$ has the value $1$;
\item
$r \in \dom(s)$ indicates that semaphore $r$ has the value $0$;
\item
$r \in \dom(s)$ and $s(r) = \emptyseq$ indicates that no process is 
suspended on \nolinebreak[2] sema\-phore $r$;
\item
if $r \in \dom(s)$ and $s(r) \neq \emptyseq$, then $s(r)$ represents 
a first-in, first-out queue of processes suspended on $r$.
\end{itemize}

As a preparation for the instantiation of the abstract schedulers 
$\sched{n}$ and control state transformers $\updat{n}$, we define some
auxiliary functions.

We define a total function $\funct{\xturns}{\Hist \x \Natpos}{\Nat}$ 
recursively as follows:
\begin{ldispl}
\begin{sgeqns}
\xturns(\emptyseq,i) = 0\;,                                          \\
\xturns(h \concat \tup{j,n},i) = 0                & \mif i \neq j\;, \\
\xturns(h \concat \tup{j,n},i) = \xturns(h,i) + 1 & \mif i = j\;.    
\end{sgeqns}
\end{ldispl}%
If $\xturns(h,i) = l$ and $l > 0$, then the interleaving history $h$ 
ends with $l$ consecutive turns of the $i$th process being interleaved.
If $\xturns(h,i)= 0$, then the interleaving history $h$ does not end 
with turns of the $i$th process being interleaved.

For each $n \in \Natpos$, we define a total function 
$\funct{\xnext{n}}{\Hist \x \Nat}{\set{1,\ldots,n}}$ by cases as 
follows:
\begin{ldispl}
\begin{sgeqns}
\xnext{n}(\emptyseq,i) = i + 1\;,                                     \\
\xnext{n}(h \concat \tup{j,n},i) = j       & \mif \xturns(h,j) < k\;, \\
\xnext{n}(h \concat \tup{j,n},i) = ((i + j) \bmod n) + 1 
                                           & \mif \xturns(h,j) \geq k\;.    
\end{sgeqns}
\end{ldispl}%
If $\xnext{n}(h,i) = j$, then the $j$th process being interleaved is the 
process that should get the $(i{+}1)$th next turn after interleaving 
history $h$ according to the round-robin scheduling algorithm, where 
each of the processes being interleaved is given $k$ consecutive turns. 
 
We define a total function $\funct{\xwaiting}{S}{\setof{\Natpos}}$ as 
follows:
\begin{ldispl}
\begin{geqns}
\xwaiting(s) = \Union{r \in \dom(s)} \elems(s(r))\;. 
\end{geqns}
\end{ldispl}%
If $\xwaiting(s) = I$, then $i \in I$ iff the $i$th process being 
interleaved is suspended on one or more semaphores in control state $s$.

For each $n \in \Natpos$, we define a partial function 
$\pfunct{\xsched{n}}{\Hist \x S \x \Nat}{\set{1,\ldots,n}}$ recursively 
as follows:
\begin{ldispl}
\begin{sgeqns}
\xsched{n}(h,s,i) = \xnext{n}(h,i)
      & \mif i < k \mul n \Land \xnext{n}(h,i) \notin \xwaiting(s)\;, \\
\xsched{n}(h,s,i) = \xsched{n}(h,s,i + 1) 
      & \mif i < k \mul n \Land \xnext{n}(h,i) \in \xwaiting(s)\;.    
\end{sgeqns}
\end{ldispl}%
The function $\xsched{n}$ is like the function $\xnext{n}$, but skips 
the processes that are suspended on one or more semaphores according to 
the control state $s$.
Notice that $\xsched{n}(h,s,i)$ is undefined if 
$\xwaiting(s) = \set{1,\ldots,n}$.
In this case, none of the processes being interleaved can be given a 
turn and the whole becomes inactive.

We define a total function 
$\funct{\xremove{n}}{S \x \set{1,\ldots,n}}{S}$ 
recursively as follows:%
\footnote
{The special function notation used below is explained in 
 Appendix~\ref{appendix-notations}.}
\begin{ldispl}
\begin{geqns}
\xremove{n}(\emptymap,i) = \emptymap\;, \\
\xremove{n}(s \owr \maplet{r}{q},i) = 
\xremove{n}(s,i) \owr \maplet{r}{\xremovep{n}(q,i)}\;,
\end{geqns}
\end{ldispl}%
where the total function 
$\funct{\xremovep{n}}
  {\seqof{\Natpos} \x \set{1,\ldots,n}}{\seqof{\Natpos}}$
is recursively defined as follows:
\begin{ldispl}
\begin{sgeqns}
\xremovep{n}(\emptyseq,i) = \emptyseq\;, \\
\xremovep{n}(j \concat q,i) = j \concat \xremovep{n}(q,i)
                                                    & \mif j < i\;, \\
\xremovep{n}(j \concat q,i) = \xremovep{n}(q,i)     & \mif j = i\;, \\
\xremovep{n}(j \concat q,i) = (j - 1) \concat \xremovep{n}(q) 
                                                    & \mif j > i\;.
\end{sgeqns}
\end{ldispl}%
If $\xremove{n}(s,i) = s'$, then $s'$ is $s$ adapted to the successful
termination of the $i$th process of the processes being interleaved.

For each $n \in \Natpos$, we instantiate the abstract scheduler 
$\sched{n}$ and control state transformer $\updat{n}$ as follows:
\begin{ldispl}
\begin{sgeqns}
\sched{n}(h,s) = \xsched{n}(h,s,0)\;,
\eqnsep
\updat{n}(\emptyseq,s,i,a) = \emptymap     & \mif a \notin C\;,       \\
\updat{n}(h \concat \tup{j,n},s,i,a) = s   & \mif a \notin C\;,       \\
\updat{n}(\emptyseq,s,i,\Pop(r)) = \maplet{r}{\emptyseq}\;,           \\
\updat{n}(h \concat \tup{j,n},s,i,\Pop(r)) = 
s \owr \maplet{r}{\emptyseq}               & \mif r \notin \dom(s)\;, \\
\updat{n}(h \concat \tup{j,n},s,i,\Pop(r)) = 
s \owr \maplet{r}{s(r) \concat i}          & \mif r \in \dom(s)\;,    \\
\updat{n}(\emptyseq,s,i,\Vop(r)) = \emptymap\;,                       \\
\updat{n}(h \concat \tup{j,n},s,i,\Vop(r)) = s             
                                           & \mif r \notin \dom(s)\;, \\
\updat{n}(h \concat \tup{j,n},s,i,\Vop(r)) = s \dsub \set{r}            
                    & \mif r \in \dom(s) \Land s(r) = \emptyseq\;,    \\
\updat{n}(h \concat \tup{j,n},s,i,\Vop(r)) = 
s \owr \maplet{r}{\tl(s(r))}             
                    & \mif r \in \dom(s) \Land s(r) \neq \emptyseq\;, \\
\updat{n}(h,s,i,\ep) = \xremove{n}(s,i)\;.
\end{sgeqns}
\end{ldispl}%
The following clarifies the connection between the instantiated control 
state transformers $\updat{n}$ and the semaphore mechanism as introduced 
in~\cite{Dij68a}:
\begin{itemize}
\item
$s = \emptymap$ indicates that all semaphores have value $1$;
\item
if $r \notin \dom(s)$, then the transition from $s$ to 
$s \owr \maplet{r}{\emptyseq}$ indicates that the value of semaphore $r$ 
changes from $1$ to $0$;
\item
if $r \in \dom(s)$, then the transition from $s$ to 
$s \owr \maplet{r}{s(r) \concat i}$ indicates that the $i$th process 
being interleaved is added to the queue of suspended processes;
\item
if $r \notin \dom(s)$, then the transition from $s$ to $s$ indicates 
that the value of semaphore $r$ remains $1$;
\item
if $r \in \dom(s)$ and $s(r) = \emptyseq$, then the transition from $s$ 
to $s \dsub \set{r}$ indicates that the value of semaphore $r$ changes 
from $0$ to $1$;
\item
if $r \in \dom(s)$ and $s(r) \neq \emptyseq$, then the transition from 
$s$ to $s \owr \maplet{r}{\tl(s(r))}$ \linebreak[2] indicates that the 
first process in the queue of suspended processes is removed from that
queue.
\end{itemize}

All respects in which the generic interleaving strategy of \siACP\ are 
widened appear to be indispensable for the instantiation presented in 
this section. 

\section{Concluding Remarks}
\label{sect-concl}

In a previous paper, we have extended the algebraic theory of processes 
known as \ACP\ with strategic interleaving, i.e.\ interleaving according
to some process-scheduling policy.
The extension concerned is based on a generic interleaving strategy that 
can be instantiated with different specific interleaving strategies.
In the current paper, we have extended the variant of \ACP\ known as 
\ACPe\ with strategic interleaving and widened the generic interleaving 
strategy in three respects.
For the widening in one of these respects, the setting of \ACP\ is 
unfit.
We have instantiated the widened generic interleaving strategy with a 
specific interleaving strategy that supports mutual exclusion of 
critical subprocesses of the different processes being interleaved.
This instantiation provides evidence of the desirability of the widening 
of the generic interleaving strategy. 

\appendix

\section{Structural Operational Semantics of \siACPe}
\label{appendix-SOS}

It is customary to associate transition systems with closed terms of the 
language of an ACP-like algebraic theory about processes by means of 
structural operational semantics and to use this to construct a model in 
which closed terms are identified if their associated transition systems 
are bisimilar.
The structural operational semantics of \ACPe\ can be found 
in~\cite{BW90}.
The additional transition rules for the strategic interleaving operators 
and the positional strategic interleaving operators are given in 
Table~\ref{trules-SI}.
\begin{table}[!t]
\caption{Transition rules for strategic interleaving}
\label{trules-SI}
\begin{ruletbl}
\RuleC
{\sterm{x}}
{\sterm{\si{1}{h}{s}{x}}}
{\sched{1}(h,s) = 1}
\\
\RuleC
{\sterm{x_i},\;
 \sterm{\si{n}{h \concat \tup{i,n}}{\updat{n+1}(h,s,i,\ep)}
        {x_1,\ldots,x_{i-1},x_{i+1},\ldots,x_{n+1}}}}
{\sterm{\si{n+1}{h}{s}{x_1,\ldots,x_{n+1}}}}
{\sched{n}(h,s) = i}
\\
\RuleC
{\astep{x_i}{a}{x'_i}}
{\astep{\si{n}{h}{s}{x_1,\ldots,x_n}}{a}
 {\si{n}{h \concat \tup{i,n}}{\updat{n}(h,s,i,a)}
  {x_1,\ldots,x_{i-1},x_i',x_{i+1},\ldots,x_n}}}
{\sched{n}(h,s) = i,\; a \notin C}
\\
\RuleC
{\astep{x_i}{a}{x'_i}}
{\astep{\si{n}{h}{s}{x_1,\ldots,x_n}}{\ol{a}}
 {\si{n}{h \concat \tup{i,n}}{\updat{n}(h,s,i,a)}
  {x_1,\ldots,x_{i-1},x_i',x_{i+1},\ldots,x_n}}}
{\sched{n}(h,s) = i,\; a \in C}
\\
\RuleC
{\astep{x_i}{\pcr(d)}{x'_i}}
{\astep{\si{n}{h}{s}{x_1,\ldots,x_n}}{\rcr(d)}
 {\si{n+1}{h \concat \tup{i,n+1}}{\updat{n}(h,s,i,\pcr(d))}
 {x_1,\ldots,x_{i-1},x_i',x_{i+1},\ldots,x_n,\crea(d)}}}
{\sched{n}(h,s) = i}
\\
\RuleC
{\sterm{x_i},\;
 \astep{\si{n}{h \concat \tup{i,n}}{\updat{n+1}(h,s,i,\ep)}
        {x_1,\ldots,x_{i-1},x_{i+1},\ldots,x_{n+1}}}{\alpha}{x'}}
{\astep{\si{n+1}{h}{s}{x_1,\ldots,x_{n+1}}}{\alpha}{x'}}
{\sched{n}(h,s) = i}
% \end{ruletbl} \end{table}
\\
% \begin{table}[!t] \caption{Transition rules for the positional
% strategic interleaving operators} \label{trules-SI-2} \begin{ruletbl}
\Rule
{\sterm{x}}
{\sterm{\posm{1}{i}{h}{s}{x}}}
\\
\Rule
{\sterm{x_i},\;
 \sterm{\posm{n}{i}{h \concat \tup{i,n}}{\updat{n+1}(h,s,i,\ep)}
        {x_1,\ldots,x_{i-1},x_{i+1},\ldots,x_{n+1}}}}
{\sterm{\posm{n+1}{i}{h}{s}{x_1,\ldots,x_{n+1}}}}
\\
\RuleC
{\astep{x_i}{a}{x'_i}}
{\astep{\posm{n}{i}{h}{s}{x_1,\ldots,x_n}}{a}
 {\si{n}{h \concat \tup{i,n}}{\updat{n}(h,s,i,a)}
  {x_1,\ldots,x_{i-1},x_i',x_{i+1},\ldots,x_n}}}
{a \notin C}
\\
\RuleC
{\astep{x_i}{a}{x'_i}}
{\astep{\posm{n}{i}{h}{s}{x_1,\ldots,x_n}}{\ol{a}}
 {\si{n}{h \concat \tup{i,n}}{\updat{n}(h,s,i,a)}
  {x_1,\ldots,x_{i-1},x_i',x_{i+1},\ldots,x_n}}}
{a \in C}
\\
\Rule
{\astep{x_i}{\pcr(d)}{x'_i}}
{\astep{\posm{n}{i}{h}{s}{x_1,\ldots,x_n}}{\rcr(d)}
 {\si{n+1}{h \concat \tup{i,n+1}}{\updat{n}(h,s,i,\pcr(d))}
 {x_1,\ldots,x_{i-1},x_i',x_{i+1},\ldots,x_n,\crea(d)}}}
\\
\Rule
{\sterm{x_i},\;
 \astep{\posm{n}{i}{h \concat \tup{i,n}}{\updat{n+1}(h,s,i,\ep)}
        {x_1,\ldots,x_{i-1},x_{i+1},\ldots,x_{n+1}}}{\alpha}{x'}}
{\astep{\posm{n+1}{i}{h}{s}{x_1,\ldots,x_{n+1}}}{\alpha}{x'}}
\end{ruletbl}
\end{table}
In this table, 
$n$ and $i$ stand for arbitrary numbers from $\Natpos$ with $i \leq n$, 
$h$ stands for an arbitrary interleaving history from $\Hist$, 
$s$ stands for an arbitrary control state from $S$, 
$a$ stands for an arbitrary action constant that is not of the form 
$\pcr(d)$ or $\rcr(d)$, 
$\alpha$ stands for an arbitrary action constant, and 
$d$ stands for an arbitrary datum $d$ from $D$.
The intuition concerning $\,\sterm{}$ and $\astep{}{a}{}$ is as follows:
\begin{itemize}
\item
$\sterm{t}$ indicates that $t$ is capable of terminating successfully;
\item
$\astep{t}{a}{t'}$ indicates that $t$ is capable of performing action 
$a$ and then proceeding as $t'$.
\end{itemize}
The transition rules for the strategic interleaving operators are 
similar to the transition rules for the positional strategic 
interleaving operators.
However, each transition rule for the strategic interleaving operators 
has the side-condition $\sched{n}(h,s) = i$.

\section{Sequence Notation and Function Notation}
\label{appendix-notations}

We use the following sequence notation:
\begin{itemize}
\item
$\emptyseq$ for the empty sequence;
\item 
$d$ for the sequence having $d$ as sole element;
\item
$u \concat v$ for the concatenation of sequences $u$ and $v$;
\item
$\hd(u)$ for the first element of non-empty sequence $u$;
\item
$\tl(u)$ for the subsequence of non-empty sequence $u$ whose first 
element is the second element of $u$ and whose last element is the last 
element of $u$;
\item
$\elems(u)$ is the set of all elements of sequence $u$. 
\end{itemize}
\pagebreak[2]
We use the following special function notation:
\begin{itemize}
\item
$\emptymap$ for the empty function; 
\item
$\maplet{d}{e}$ for the function $f$ with $\dom(f) = \set{d}$ such that
$f(d) = e$; 
\item
$f \owr g$ for the function $h$ with $\dom(h) = \dom(f) \union \dom(g)$
such that for all $d \in \dom(h)$, $h(d) = f(d)$ if
$d \notin \dom(g)$ and $h(d) = g(d)$ otherwise;
\item
$f \dsub S$ for the function $g$ with $\dom(g) = \dom(f) \diff S$
such that for all $d \in \dom(g)$, $g(d) = f(d)$.
\end{itemize}

\bibliographystyle{splncs03}
\bibliography{PA}

\begin{comment}
A scheduling policy that deals with mutual exclusion locks for 
controlling access to shared resources is represented in the way 
assumed by the theory developed to demonstrate the suitability of this
way of representation.

In the work presented in this paper, we consider strategic interleaving
where process creation is taken into account.
The approach to process creation followed in this paper originates from
the one first followed in~\cite{Ber90a} to extend \ACP\ with process 
creation and later followed \linebreak[2] in~\cite{BB93a,BMU98a,BM02a} 
to extend different timed versions of \ACP\ with process creation.
The only other approach that we know of is the approach, based 
on~\cite{AB88a}, that has for instance been followed 
in~\cite{BV92a,GR97a}.
However, with that approach, it is most unlikely that data about the
creation of processes can be made available for the decision making 
con\-cerning the strategic interleaving of processes.
\end{comment}

\end{document}